\documentclass[12pt]{article}
\usepackage{amssymb,amsmath,epsfig}
\allowdisplaybreaks

\begin{document}

\title{\bf Dynamics of Modified Chaplygin Gas Inflation on the Brane with Bulk Viscous Pressure}
\author{Abdul Jawad \thanks{jawadab181@yahoo.com;~~abduljawad@ciitlahore.edu.pk}, Amara
Ilyas\thanks{amara$\_$Ilyas14@yahoo.com}~ and Shamaila Rani\thanks{drshamailarani@ciitlahore.edu.pk } \\
Department of Mathematics, COMSATS Institute of\\ Information
Technology, Lahore-54000, Pakistan.}

\date{}
\maketitle
\begin{abstract}
We investigate the role of bulk viscous pressure on the warm
inflationary modified Chaplygin gas in brane-world framework in the
presence of standard scalar field. We assume the intermediate
inflationary scenario in strong dissipative regime and constructed
the inflaton, potential, entropy density, slow-roll parameters,
scalar and tensor power spectra, scalar spectral index and
tensor-to-scalar ratio. We develop various trajectories such as $n_s
- N$, $n_s - r$ and $n_s - \alpha_s$ (where $n_s$ is the spectral
index, $\alpha_s$ is the running of spectral index, $N$ is the
number of e-folds and $r$ is tensor-to-scalar ratio) for variable as
well as constant dissipation and bulk viscous coefficients at high
dissipative regime. It is interesting to remark here that our
results of these parameters are compatible with recent observational
data such as WMAP $7+9$, BICEP$2$ and Planck data.
\end{abstract}
Chaplygin gas model
; Inflationary parameters.\\

\section{Introduction}

At early times, there was a segment in which the universe evolved
through accelerated expansion in a short period of time at high
energy scales which result the idea of inflation. Inflationary
theory has major achievements in solving the longstanding
cosmological puzzles like monopole, flatness, horizon etc
\cite{a1,a2}. The mechanism of large-scale structure (LSS) and
anisotropy of cosmological microwave background (CMB) is explained
by inflationary theory \cite{a3,a4}. In warm inflationary models,
the radiation production arises during inflationary period and
reheating can be avoided \cite{w3}. The thermal fluctuations could
be obtain in these models which produce initial fluctuations and are
crucial for LSS formation. Warm inflationary period ends when the
universe stops inflating and after that, the universe enters in
radiation phase smoothly \cite{w3}. In the end, remaining inflatons
or dominant radiation fields produced the matter components of the
universe. For the sake of simplicity, the particles (which are
produced due to inflaton decay) are assumed as massless particles
(or radiation) in warm inflation models.

The existence of massive particles has been considered in \cite{w5}
and corresponding perturbation parameters of this model have been
presented in \cite{w6}. Decay of the massive particles within the
fluid is an entropy-producing scalar phenomenon, on the other hand
"bulk viscous pressure" has entropy-producing property. Therefore
the decay of particles may be assumed as a bulk viscous pressure
$\Pi = -3\xi H$ \cite{w7}, where $H$ and $\xi$ are Hubble parameter
and phenomenological coefficient of bulk viscosity, respectively.
This coefficient is positive-definite by the second law of
thermodynamics and depends on the energy density of the fluid. The
inflationary epoch can be divided into epochs such as slow-roll and
reheating. During the slow-roll approximation, the universe inflates
as the interactions between inflatons and other fields become
negligibly small and the potential energy dominates the kinetic
energy. After this period, the universe enters the last stage of
inflation, i.e., the reheating era, in which the kinetic and
potential energies are comparable. Here the inflation starts to
oscillate around the minimum of its potential while losing its
energy to massless particles. During inflationary phase, the forms
of energy density like radiation or matter were dominated by the
vacuum energy while the scale factor increased exponentially over
time \cite{B2}.

The cosmic acceleration in the early universe (or inflationary
universe) can be realized by a scalar field (inflaton) through an
effective potential which represents the evolution of this field.
The scalar field models deal with two pictures of inflation, i.e.,
slow-roll and reheating. During the slow-roll phase, the universe
undergoes a rapid expansion while kinetic term of inflaton is less
than that of potential energy. In the reheating period, these two
energies are comparable and inflaton starts to oscillate about the
minimum of the potential by loosing its energy to other radiation
fields \cite{B9}. Moreover,  MCG explains the expansion of the
universe from phase dominated for small values of cosmological scale
factor using EoS i.e, $p= \omega \rho$ to large values of scale
factor using cosmological constant i.e,
$(\frac{\beta}{1+\omega})^\frac{1}{1+n}$ \cite{a}. A fluid of
(modified) CG is usually applied to explain the late time
acceleration of our universe as a possible candidate of dark energy.
MCG mimics the behavior of matter at early-times and that of a
cosmological constant at late-times. CMB also indicated that the
early universe also passes through an accelerating phase too, which
is the inflationary epoch. Given the attractiveness of the MCG as a
dark energy candidate, a natural question to ask is: Can inflation
be accommodated within the MCG scenario? This is the question we
wish to address in the present work. However, we should emphasize
that our inflationary model is not presented as a more desirable
alternative to the conventional ones. Rather, we merely aim to
establish the assumptions and extrapolations required to obtain
successful inflation in a Chaplygin inspired model \cite{B6}.

A feasible solution called intermediate inflation scenario exists in
the literature which is defined as $a = \exp(At^{f})$, where $f$ is
constant ($0 < f < 1$). In this scenario, the expansion of the
universe is slower than standard de Sitter inflation $(a = e^{Ht})$
while faster than power law inflation $(a =t^{p},~ p > 1)$. This
method was introduced for a particular scalar field potential of the
type $V (\phi) \propto \phi ^{-4(f^{-1}-1)}$ as the universe was
controlled by a potential $V(\phi)$ during inflationary period
\cite{a20, a21}. In the slow-roll estimation with this type of
potential, it is possible to have a spectrum of density
perturbations which presents a scale invariant spectral index, i.e.
$n_{s}=1$.

Yokoyama and Maeda \cite{a22} worked on intermediate inflation in
the brane-world scenario and demonstrated the nonzero value of
tensor-scalar ratio. Bamba et al. \cite{bam1} have considered
inflationary cosmology in a viscous fluid model. Setare and Kamali
\cite{a24} investigated warm-viscous inflation on the brane by
taking chaotic potential and found that the values of all involved
parameters are consistent with Wilkinson Microwave Anisotropy Probe
(WMAP)$9$, Planck and BICEP$2$ observational data. Herrera and Campo
\cite{a25} examined the parameters slow-roll parameters, inflaton,
energy density, entropy density, etc on the brane intermediate
inflationary model in high dissipative regime and found the
consistency with WMAP$5$. Setare and Kamali \cite{a26} worked on the
brane with warm-viscous inflationary universe model and tachyon
scalar field. They have calculated the parameters of this
inflationary model which are compatible with WMAP$7$. Setare and
Kamali \cite{a27} studied the tachyon-warm intermediate and
logamediate inflation in the brane-world scenario by taking
exponential potential and compared their results with recent
observational data from WMAP$9$ and Planck data.

In the present work, we investigate the inflationary parameters for
warm intermediate inflation with bulk viscous pressure in high
dissipative scenario. The outline of the paper is as follows: Basic
equations related to braneworld model and Chaplygin gas (CG) models
are discussed in section \textbf{2}. In section \textbf{3}, we will
consider different characteristics of warm intermediate inflation
and examine the results of slow$-$roll parameters according to
braneworld model along with modified Chaplygin gas (MCG). Detailed
discussions of perturbed parameters for variable coefficients as
well as constant coefficients will take place in section \textbf{4}.
Section \textbf{5} contains the conclusion.

\section{Braneworld Model}

In the braneworld scheme, the observable four-dimensional universe
is considered as a domain wall implanted on a higher-dimensional
bulk space. Einstein's field equation in the braneworld theory with
cosmological constant as a matter and source fields bounded to
3-brane may be designed as follows \cite{b3}
\begin{equation}\nonumber
G_{\mu\nu} = -\Lambda_4 g_{\mu\nu}
+\bigg(\frac{8\pi}{M_4^2}\bigg)T_{\mu\nu} +
\bigg(\frac{8\pi}{M_5^2}\bigg)^{2} \pi_{\mu\nu} - E_{\mu\nu},
\end{equation}
where $E_{\mu\nu}$ is a projection of $5$d Weyl tensor, $M_4$ and
$M_5$ are Planck scales in $4$ and $5$ dimensions respectively,
$T_{\mu\nu}$ is energy density tensor on the brane and
$\pi_{\mu\nu}$ is a tensor quadratic in $T_{\mu\nu}$. Effectual
cosmological constant $\Lambda_4$ on the brane in terms of 3-brane
tension $\sigma$ and cosmological constant ($\Lambda$) is given by
\begin{equation}\nonumber
\Lambda_4 = \frac{4\pi}{M_5^3}\bigg(\Lambda +
\frac{4\pi}{3M_5^3}\sigma^2\bigg),
\end{equation}
and $4$d Planck scale is determined by $5$d Planck scale as
\begin{equation}\nonumber
M_4 =
\bigg(\frac{3}{4\pi}\bigg)^\frac{1}{2}\bigg(\frac{M_5^2}{\sqrt{\sigma}}\bigg)M_5.
\end{equation}

The spatially flat FRW space comprised by line element as follows
\begin{equation}\nonumber
ds^{2}= -dt^{2}+a^{2}(t)(dx^{2}+dy^{2}+dz^{2})
\end{equation}
where $a(t)$ is the scale factor. For flat FRW model, Friedmann
equation on the brane turns out to be \cite{b3}
\begin{equation}\nonumber
H^2 = \frac{\Lambda_4}{3} +\bigg(\frac{8\pi}{3 M_4^2}\bigg)\rho_\tau
+ \bigg(\frac{4\pi}{3 M_5^2}\bigg)\rho_\tau^2 +
\frac{\varepsilon}{a^4},
\end{equation}
where $\rho_\tau$ is total energy density on the brane. In the above
equation, last term denotes the shape of the bulk gravitons on the
brane, where $\varepsilon$ is an integration constant which comes up
from Weyl tensor $E_{\mu\nu}$. The projected Weyl tensor term in the
effectual Einstein equation may be neglected because this term may
be speedily stretched when the inflation starts \cite{a26}. It is
also assumed that the $\Lambda_4$ is negligible in the early
universe. The Friedmann equation is reduced to
\begin{equation}\nonumber
H^2= \frac{8\pi}{3 M_4^2}\rho_\tau\bigg(1+\frac{\rho_\tau}{2
\lambda}\bigg).
\end{equation}

The equation of state of pure CG is defined
as
\begin{equation*}
p=-\frac{\beta}{\rho},
\end{equation*}
where $\beta$ is positive parameter, $p$ represents pressure and
$\rho$ is the energy density. The extended form of CG was driven by
Kamenshchik et al. \cite{a18}, named as generalized CG (GCG) is
given by
\begin{equation*}
p = -\frac{ \beta}{{\rho}^ \sigma}, \quad 0 < \sigma < 1.
\end{equation*}
An advance study of CG called MCG has been introduced by Benaoum
\cite{bam2,a19} with the following equation of state
\begin{equation*}
p_{MCG} = \omega\rho_{MCG} - \frac{\beta}{{\rho_{MCG}}^ \sigma},
\quad 0 < \sigma < 1
\end{equation*}
where $\omega$ is a positive constant. The energy conservation
equation for MCG model described as:
\begin{equation}\nonumber
\rho_{MCG}=\bigg(\frac{\beta}{1+\omega}+\frac{\upsilon}{a^{3(1+
\sigma)(1+\omega)}}\bigg)^{\frac{1}{1+\sigma}},
\end{equation}
where $\upsilon$ is constant of integration.

\section{Basic Inflationary Scenario}

Now we study different characteristics of intermediate inflationary
MCG model for FRW universe distinguished by inflaton and matter
radiation. By assuming perturbed parameters, we fix our model
everywhere at intermediate epoch and assure the compatibility with
observational data. The warm MCG model including imperfect fluid on
the brane modifies first Friedmann equation as \cite{aa25}
\begin{eqnarray}\nonumber
H^{2}&=&\frac{8\pi}{3M_{4}^{2}}\bigg(\rho_\phi+\rho\bigg)\bigg(1
+\frac{\rho_\phi+\rho}{2\lambda}\bigg)\\\nonumber
&=&\frac{8\pi}{6M_{4}^{2}\lambda}\bigg[\bigg(\frac{\beta}{1+\omega}
+\upsilon\rho_{\phi}^{(1+\sigma)(1+\omega)}\bigg)^{\frac{1}{1+\sigma}}+T
S(\phi,T)\bigg]\bigg[2\lambda+\bigg(\frac{\beta}{1+\omega}\\\label{f}
&+&\upsilon\rho_{\phi}^{(1+\sigma)(1+\omega)}\bigg)^{\frac{1}{1+\sigma}}+T
S(\phi,T)\bigg],
\end{eqnarray}
where we have used $\rho_\tau=\rho_\phi+\rho$ and $\rho$ is taking
as energy density in terms of imperfect fluid $\rho= TS(\phi,T)$
($T$ represents temperature and $S$ stands for entropy density).
Also, $\rho_\phi$ is assumed to be the energy density of MCG.

Since $\rho_\phi \ll \rho_\phi^2 $, then Eq.(\ref{f}) leads to
\cite{a25}
\begin{eqnarray}\nonumber
H^{2}&=&\frac{8\pi}{6M_{4}^{2}\lambda}\bigg[\bigg(\frac{\beta}{1+\omega}
+\upsilon\rho_{\phi}^{(1+\sigma)(1+\omega)}\bigg)^{\frac{1}{1+\sigma}}+T
S(\phi,T)\bigg]^2.
\end{eqnarray}
The most important parameters of inflation are energy density
$\rho_{\phi}$ and pressure $p_{\phi}$ which can be defined as
follows
\begin{equation}\nonumber
\rho_{\phi}=\frac{\dot{\phi}^{2}}{2}+V(\phi),\quad
p_{\phi}=\frac{\dot{\phi}^{2}}{2}-V(\phi),
\end{equation}
where dot denotes derivative with respect to time $t$ and  $V(\phi)$
is used as potential term.

The inflaton and imperfect fluid energy densities are conserved as
\begin{equation}\label{r}
\dot{\rho}+3H(\psi\rho+\Pi)=\Gamma\dot{\phi}^{2}.
\end{equation}
\begin{equation}\label{rr}
\dot{\rho_\phi} +3H(\rho_\phi+p_\phi)=-\Gamma\dot{\phi}^{2}.
\end{equation}
Here, we have used the bulk viscous pressure defined as as $p +
\Pi$, while $p=(\psi-1)\rho$ with adiabatic index $1\leq \psi \leq
2$ and $\Pi=-3 \xi H$ is bulk viscous pressure ($\xi$ is the
coefficient of bulk viscosity which is positive-definite and
generally depends on $\rho$). Also $\Gamma$ is the dissipation
factor that assess the rate of decay of $\rho_\phi$ into $\rho$. The
second law of thermodynamics indicates that $\Gamma$ must be
positive, so the inflaton energy density decompose into radiation
density. Energy density of imperfect fluid is $\rho=TS(\phi,T)$
which changes the Eq.(\ref{r}) as
\begin{equation*}
S\dot{T}+T\dot{S}+3H(\psi TS+\Pi)=\Gamma\dot{\phi}^{2},
\end{equation*}
here, $\dot{T}$ is negligible. The negativity of bulk viscous
pressure contributes to increase the source of entropy density given
on the right-hand side of the above equation. The second
conservation equation can be written in view of $\rho_\phi$ and
$p_\phi$ as
\begin{equation*}
\ddot{\phi}+(3H+\Gamma)\dot{\phi}+\acute{V(\phi)}=0\Rightarrow\ddot{\phi}
+3H(1+\frac{\Gamma}{3H})\dot{\phi}+\acute{V(\phi)}=0
\end{equation*}
where $\frac{\Gamma}{3H}=R$ and prime represents the derivative with
respect to $\phi$. In weak dissipative epoch, $R\ll1$ runs to
$\Gamma\ll3H$ while $R\gg1$ denotes high dissipative era (where
dissipation coefficient is much bigger than Hubble scale). The
evolution of the universe during inflationary regime diffuses the
decay of inflaton for $R\ll1$. Bulk viscosity cannot be neglected
when the sectors of the inflationary fluid interact with each other
and turns negligible throughout this region. Implementation of
limits required for getting the epoch to be static, i.e.
$\rho_\phi\approx V(\phi),\rho < \rho_\phi$, slow-roll limit,
$V(\phi)\gg \dot{\phi}^{2}$, $(3H+\Gamma)\dot{\phi}\gg\ddot{\phi}$
and quasi-stable decay of $\rho_\phi$ into $\rho$, where
$3H(\psi\rho+\Pi)\gg \dot{\rho}$, $\Gamma\dot{\phi}^{2}\gg
\dot{\rho}$. When all these limits apply in the Eqs.(\ref{r}) and
(\ref{rr}) then dynamical equations takes the form as
\begin{eqnarray}\label{b}
H^{2}&=&\frac{8\pi}{6M_{4}^{2}\lambda}\bigg(\frac{\beta}{1+\omega}
+\upsilon\rho_{\phi}^{(1+\sigma)(1+\omega)}\bigg)^{\frac{2}{1+\sigma}},\\\label{v}
\dot{\phi}&=&\frac{-V'(\phi)}{3H(1+r)},\\\label{vv}
\psi\rho&=&r\dot{\phi}^{2}-\Pi.
\end{eqnarray}
The essential slow-roll approximation is controlled by a set of
dimensionless slow-roll parameters which are defined as \cite{b8}
\begin{equation*}
\epsilon = -\frac{\dot{H}}{H^2},~~~~~~~~~~~~ \eta =
-\frac{\ddot{H}}{H\dot{H}}.
\end{equation*}
These parameters for inflationary viscous universe model can be
represented as
\begin{eqnarray}\label{h}
\epsilon &=&\left(\frac{3\lambda  M_4^2}{4 \pi}\right)^{2}
\frac{\upsilon (1+\omega )  V^{-1+(1+\sigma ) (1+\omega )}V'^2}{R
\left(\frac{\beta }{1+\omega }+\upsilon V^{(1+\sigma ) (1+\omega
)}\right)^{1+\frac{4}{1+\sigma }}}.
\\\nonumber\eta &=& 3 \lambda  M_4^2  \left(\beta
(\sigma +\omega +\sigma  \omega )-\upsilon(1+\omega ) V^{(1+\sigma )
(1+\omega )}\right) V'^2+2V\bigg(\beta +\upsilon  (1+\omega
)\\\nonumber&&\times V^{(1+\sigma ) (1+\omega )}\bigg) V''\bigg[12
\pi (1+R) V \left(\beta +\upsilon (1+\omega )V^{(1+\sigma )
(1+\omega )}\right)\bigg(\frac{\beta }{1+\omega
}+\upsilon\\\label{hh}&&\times V^{(1+\sigma ) (1+\omega
)}\bigg)^{\frac{2}{1+\sigma }} \bigg]^{-1}.
\end{eqnarray}
These parameters are smaller than $1+R$ lead to the sensible
slow-roll limit. The condition $\epsilon < 1$ leads to $\rho_\phi
> 3(1+\omega)(\psi\rho+\Pi)^{1-(1+\sigma)(1+\omega)}$. The slow-roll
inflation ends at $\epsilon = 1$. The inflation not only demands
$\epsilon < 1$ but also $\eta$ must be small over a reasonably large
period of time, enough number of e-folds represented by $N$. We
bound our model in the region $\phi > 0$. This number can be
calculated by using the following formula
\begin{eqnarray}\label{hhh}
N(\phi)&=& \int^{t_g}_{t_i} H dt = \int^{\phi_g}_{\phi_i}(4\pi
R)(V'\lambda M_4^2)^{-1}\bigg(\frac{\beta}{1+\omega}+\upsilon
V^{(1+\sigma)(1+\omega)}\bigg)^{\frac{2}{1+\sigma}}
\end{eqnarray}
where $\phi_i$ and $\phi_g$ stand for initial and final inflatons
respectively.

Also, Eq.(\ref{vv}) and slow-roll parameter ($\epsilon$) provides
the following relationship between $\rho_\phi$ and $\rho$ including
the effect of viscous pressure as
\begin{eqnarray}\nonumber
\psi\rho+\Pi&=&\bigg[\epsilon\bigg(2\lambda+\bigg(\frac{\beta}{1+
\omega}+\upsilon\rho_\phi^{(1+\sigma)(1+\omega)}\bigg)^\frac{1}
{1+\sigma}\bigg)\bigg(\frac{\beta}{1+\omega}+\upsilon\rho_\phi^{(1
+\sigma)(1+\omega)}\bigg)\\\nonumber&\times&\rho_\phi^{1-(1+\sigma)
(1+\omega)}\bigg]\bigg[3\upsilon(1+\omega)\bigg(\lambda+\bigg(\frac
{\beta}{1+\omega}+\upsilon\rho_\phi^{(1+\sigma)(1+\omega)}\bigg)^\frac{1}
{1+\sigma}\bigg)\bigg]^{-1}.
\end{eqnarray}

\section{Perturbations}

We consider different types of perturbations for FRW background and
measure tensor and scalar disorders at minor stage by changing the
value of $\phi$. There are basically four quantities which are
mostly analyzed for inflationary disorders, i.e., tensor and scalar
power spectra $(P_r,P_s)$, tensor and scalar spectral indices $(n_r,
n_s)$. The form of scalar power spectrum can be estimated as
$P_r(k_0)\equiv\frac{25}{4}\delta_H^2(k_0)$, where $K_F
=\sqrt{\Gamma H}$ and density disorders
$\delta_H^2(k_0)\equiv\frac{K_F (T_r)}{2\pi^2}$.
We analyze the change in value of inflationary parameters by
considering two different cases, i.e., \textbf{(i)} taking bulk and
dissipation coefficients as variables \textbf{(ii)} taking bulk and
dissipation coefficients as constants.
\begin{figure}
\centering \epsfig{file=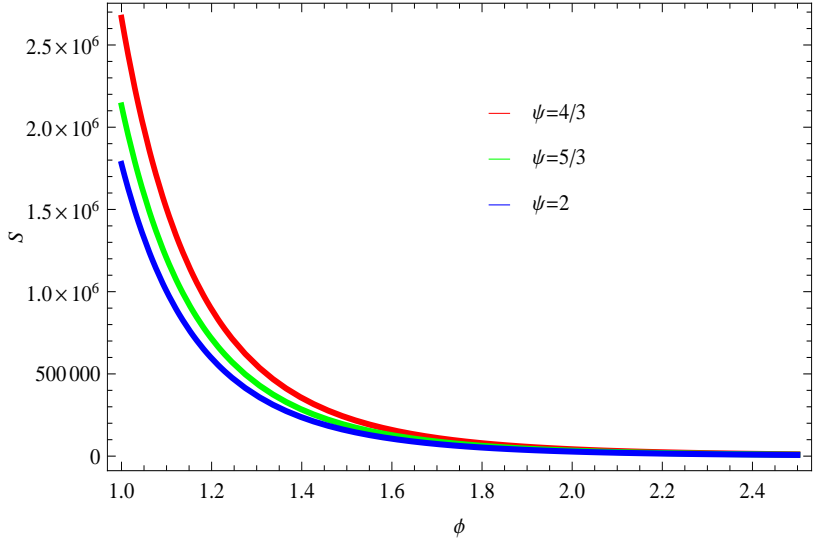, width=.80\linewidth,
height=2.9in} \caption{Plot of entropy density ($S(\phi)$) versus
scalar field for $\lambda= 1,~\sigma =1,~ A = 1,~ M_4 = 1,~ f =
\frac{3}{5},~\Gamma = 0.275,~\xi_1 = 0.2 \times 10^{-8},~T
=5.47\times 10^{-5},~\omega =0.25,~\beta = 0.775,~\upsilon = 1$.}
\end{figure}

\subsection{Variable Bulk and Dissipative Coefficients}

Here, we choose $\Gamma=\Gamma(\phi)= V(\phi)$ and
$\xi=\xi(\rho)=\xi_1\rho$. With these constraints, inflaton and
effective potential can be obtained by using Eqs.(\ref{b})-(\ref{v})
\begin{eqnarray}\nonumber
\phi -\phi _0 &=& 2 (1+b)t^{\frac{1}{2} (1+b)} ( \upsilon (1+\omega
))^{\frac{1}{2}}(f-1)^{\frac{-1}{2}}.\\\label{e} V(\phi )&=&
\frac{3(A f)^2 M_4^2\lambda }{4\pi } \bigg[( \upsilon  (1+\omega
))^{\frac{-1}{2}}(f-1)^{\frac{1}{2}}(2+2b)^{-1}\bigg]^{\frac{4(f-1)}{1+b}}\phi
^{\frac{4(f-1)}{1+b}}.
\end{eqnarray}
respectively, and $b = (f-1) (1+\sigma )$. By using
Eqs.(\ref{v})-(\ref{vv}), the entropy density leads to
\begin{eqnarray}\nonumber
S(\phi)&=&\frac{16 A f b_1^2\bigg(\frac{1}{2(1+b)}\bigg)^{2 b_1}
\bigg(\frac{f-1}{\upsilon  (1+\omega )}\bigg)^{2 b_1} \phi ^{-2+2
b_1} }{T \psi\bigg(1 -3 A \xi_1 \bigg(\frac{1}{2(1+b)}\bigg)^{2
b_1}\bigg(\frac{f-1}{\upsilon (1+\omega )}\bigg)^{2 b_1}\phi ^{2
b_1} \bigg)}.
\end{eqnarray}
where $b_1=\frac{f-1}{1+b}$. We plot $S(\phi)$ versus $\phi$ in
Figure \textbf{1} and observe that it shows the increasing behavior
of entropy density with respect to $\phi$ for three different values
of $\psi$. There is no change in present model with and without bulk
viscous pressure $\Pi$.

However, the slow roll parameter $\epsilon$ in terms of $\phi$ by
using Eq. (\ref{h}) turns out to be
\begin{eqnarray}\label{ee}
\epsilon &=&\frac{1-f}{A f}\bigg[\phi ^{\frac{2}{1+b}} ( \upsilon
(1+\omega ))^{\frac{-1}{2}}(f-1)^{\frac{1}{2}}(2+2b)^{-1}\bigg]^{-f}
\end{eqnarray}
By setting $\epsilon=1$, we can get the value of lower bound on
inflatons (defined as $\phi_i$) as follows
\begin{eqnarray}\nonumber
\phi _i&=& 2(1+b)\bigg(\frac{1-f}{A f}\bigg)^{\frac{1+b}{2f}}\bigg(
\frac{\upsilon  (1+\omega )}{f-1}\bigg)^{\frac{1}{2}}.
\end{eqnarray}
This lower bound of inflatons helps us in finding its upper bound in
terms of number of e-folds ($N$) by taking Eq. (\ref{hhh}) which is
given by
\begin{eqnarray}\label{phif}
\phi _f=(2(1+b))\bigg(\frac{\upsilon  (1+\omega
)}{f-1}\bigg)^{\frac{1}{2}}\bigg[\frac{N}{A}+\bigg(\frac{1-f}{A
f}\bigg)\bigg]^{\frac{1+b}{2f}}.
\end{eqnarray}
The scalar power spectrum ($P_r$) in high dissipative era and
amplitude of tensor perturbations $(A_g^2)$ are computed by using
the following formulae
\begin{equation}\nonumber
P_r = \frac{T_r}{2 \pi^2 \epsilon\sqrt{R
V^3}}\exp(-2\chi(\phi)),\quad A_g^2 = 2 \bigg(\frac{H}{2\pi}\bigg)^2
\coth\bigg[\frac{K}{2T}\bigg],
\end{equation}
respectively. Here $T_r$ and $T$ are temperatures of air current
fluctuations and thermal background of gravitational waves,
respectively. Stimulated emissions rise up in the thermal background
of gravitational waves during the multiplication of tensor
perturbation in inflation. Therefore, the temperature of thermal
background of gravitational waves has got an extra factor
$\coth(\frac{K}{2T})$, where $K$ is the wave number. Further,
$\chi(\phi)$ (an auxiliary function) in high dissipative regime can
be defined as follows \cite{a9}
\begin{equation}\label{d}
\chi(\phi)= -\int\bigg\{\frac{\Gamma'}{3 H
R}+\frac{3}{8}\bigg[1-\bigg((\psi-1)+\frac{\Pi}{\xi}\frac{d\xi}{d\rho}
\bigg)\frac{\Gamma'V'}{9R\psi H^2}\bigg]\frac{V'}{V}\bigg\}d\phi.
\end{equation}
By using Eqs.(\ref{b}), (\ref{e}) in Eq.(\ref{d}), $\chi$ takes the
form
\begin{eqnarray}\nonumber
\chi(\phi)&=&-\frac{11}{8}\ln\phi+\frac{3(A f) }{ 16\psi  }
\bigg(\frac{1}{2(1+b)}\bigg)^{2b_1}\bigg(\frac{f-1}{\upsilon
(1+\omega )}\bigg)^{b_1}\phi ^{2(b_1-1)}\\\nonumber&\times&
\bigg[\bigg(\frac{f-1}{\upsilon  (1+\omega
)}\bigg)^{b_1}\bigg(\frac{4 b_1(\psi  - 1)}{ b_1-1 }\bigg)
-\bigg(\frac{M_4^2A f \lambda }{4\pi }\bigg)
\bigg(\frac{1}{2(1+b)}\bigg)^{2b_1}
\\\label{w}&\times&\bigg(\frac{\xi_1 }{2b_1-1}\bigg)\phi ^{2b_1}\bigg]
\end{eqnarray}

The scalar power spectrum ($P_r$) and amplitude of tensor
perturbations $(A_g^2)$ in high dissipation epoch can be obtained by
substituting the expressions (\ref{e}), (\ref{ee}), (\ref{w}) as
\begin{eqnarray}\nonumber
P_r&=& \frac{T_r\sqrt{3A f}}{2 \pi ^2} \bigg(\frac{4\pi }{3M_4^2(A
f)^2\lambda }\bigg)^2 \bigg(\frac{A f}{1-f}\bigg)
\bigg(\frac{f-1}{\upsilon  (1+\omega )}\bigg)^{\frac{7-5f}{2(1+b)}}
\bigg(\frac{1}{2(1+b)}\bigg)^{\frac{7-5f}{1+b}}\\\nonumber&\times&\phi
^{\frac{7-5f}{1+b}} \exp\bigg[\frac{11}{4}\ln\phi-\frac{3(A f) }{
4\psi }\bigg(\frac{1}{2(1+b)}\bigg)^{2b_1}\bigg(\frac{f-1}{\upsilon
(1+\omega )}\bigg)^{b_1}\phi
^{2(b_1-1)}\\\nonumber&\times&\bigg[\bigg(\frac{f-1}{\upsilon
(1+\omega )}\bigg)^{b_1}\bigg(\frac{4 b_1(\psi  - 1)}{ b_1-1
}\bigg)-\bigg(\frac{M_4^2A f \lambda }{4\pi }\bigg)
\bigg(\frac{1}{2(1+b)}\bigg)^{2b_1}\\\nonumber&\times&\bigg(\frac{\xi_1
}{2b_1-1}\bigg)\phi ^{2b_1}\bigg]\bigg],\\\nonumber A_g^2&=&\frac{(A
f)^2 }{2 \pi ^2}\bigg(\frac{1}{2(1+b)}\bigg)^{\frac{4(f-1)}{1+b}}
\bigg(\frac{f-1}{\upsilon  (1+\omega
)}\bigg)^{\frac{4(f-1)}{1+b}}\coth \bigg[\frac{K}{
2T}\bigg]_{K=K_0}\phi ^{\frac{4(f-1)}{1+b}}.
\end{eqnarray}

Moreover, the ratios of factors $P_r$ to $A_g^2$ in high dissipative
regime called the tensor-to-scalar ratio which is given by
\begin{equation}\nonumber
r(K_0)=\frac{P_r}{A_g^2}=\frac{2}{3}\bigg[ \frac{\epsilon\sqrt{R
V^5}}{T_r}\bigg]exp(2\chi(\phi))\coth\bigg[\frac{K}{2T}\bigg]_{K=K_0}
\end{equation}
where $K_0= 0.002 Mpc^{-1}$ is the pivot point. The computation of
this ratio for the present scenario turns out to be
\begin{eqnarray}\nonumber
r(K_0)&=&\frac{(1-f)\sqrt{A
f}}{T_r\sqrt{3}}\bigg(\frac{f-1}{\upsilon (1+\omega
)}\bigg)^{\frac{13f-15}{1+b}}\bigg(\frac{1}{2(1+b)}\bigg)^{\frac{9f-11}{1+b}}\bigg(\frac{3M_4^2(A
f)^2\lambda }{4\pi }\bigg)^2\\\nonumber&\times& \coth\bigg[\frac{K}{
2T}\bigg]_{K=K_0}\phi ^{\frac{9f-11}{1+b}}
\exp\bigg[-\frac{11}{4}\ln\phi+\frac{3(A f) }{ 4\psi
}\bigg(\frac{1}{2(1+b)}\bigg)^{2b_1}\\\nonumber&\times&\bigg(\frac{f-1}{\upsilon
(1+\omega )}\bigg)^{b_1}\phi
^{2(b_1-1)}\bigg[\bigg(\frac{f-1}{\upsilon  (1+\omega
)}\bigg)^{b_1}\bigg(\frac{4 b_1(\psi  - 1)}{ b_1-1
}\bigg)\\\nonumber&-&\bigg(\frac{M_4^2A f \lambda }{4\pi }\bigg)
\bigg(\frac{1}{2(1+b)}\bigg)^{2b_1}\bigg(\frac{\xi_1
}{2b_1-1}\bigg)\phi ^{2b_1}\bigg]\bigg].
\end{eqnarray}
\begin{figure}
\centering \epsfig{file=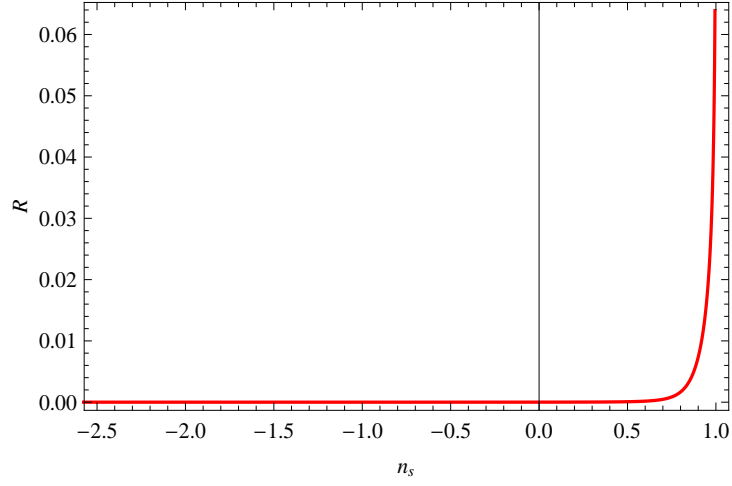, width=.70 \linewidth, height=2.5in}
\caption{Graph for scalar-tensor ratio verses spectral index for
$\lambda = 0.5, \sigma =1, A =1, M_4= 1, f = \frac{3}{7}, \psi =
\frac{5}{3}, \xi_1 = 0.2\times 10^{-8}, T = 5.47\times
10^{-5};\omega = 0.25, \upsilon = 1$.}
\end{figure}

It can be observed from Figure \textbf{2} that the scalar-tensor
ratio ($r$) remains less than $0.88$ for the range of spectral index
$0.85 < n_s < 0.98$. However, an upper bound for tensor-to-scalar
ratio as predicted by the BISEP$2$ \cite{b6}, WMAP$7$ \cite{b1, b2},
WMAP$9$ \cite{b4} and Plank data \cite{b5} are  $r <
0.26,~0.36,~0.38,\\~0.11$ respectively. Hence, our results show the
compatibility with mentioned observational data.

Moreover, the scalar spectral index $n_s$ can be defined as
\cite{b7}
\begin{equation}\nonumber
n_s - 1 =\frac{d\ln{P_r}}{d\ln{K}}.
\end{equation}
Here, wave number $K$ is related to the number of e-folds $N$
through relation $d\ln{K}= - d{N}$. With the help of above
equations, $n_s$ takes the form as
\begin{eqnarray}\nonumber
n_s&=&1+\frac{4^{b_1}}{(A f)} \bigg(\frac{1}{1+b}\bigg)^{1-2 b_1}
\exp \bigg[-b_3 b_5\phi ^{4 b_1-2} \bigg] \phi ^{-2 (2+b_1)}
\bigg(\frac{f-1}{\upsilon  (1+\omega )}\bigg)^{-b_1}
b_1\\\nonumber&\times& \bigg[(11 (1+b) \exp\bigg[b_3 b_5\phi ^{2
b_1} \bigg]-4 \exp\bigg[b_3 b_5\phi ^{4 b_1-2} \bigg] (5f-7)) \phi
^2+8\\\nonumber&\times& (1+b) \exp\bigg[b_3 b_5\phi ^{2 b_1} \bigg]
\phi ^{2 b_1} b_3 ((1-b_1) b_4+\phi ^2 b_1 b_5)\bigg],
\end{eqnarray}
where
\begin{eqnarray}\nonumber
b_2&=&\frac{\sqrt{3A f} T}{2 \pi ^2} \bigg(\frac{4\pi }{3M_4^2(A
f)^2\lambda }\bigg)^2 \bigg(\frac{A f}{1-f}\bigg)
\bigg(\frac{f-1}{\upsilon  (1+\omega )}\bigg)^{\frac{7-5f}{2(1+b)}}
\bigg(\frac{1}{2(1+b)}\bigg)^{\frac{7-5f}{1+b}},\\\nonumber
b_3&=&\frac{3(A f) }{ 4\psi
}\bigg(\frac{1}{2(1+b)}\bigg)^{2b_1}\bigg(\frac{f-1}{\upsilon
(1+\omega )}\bigg)^{b_1},\\\nonumber b_4&=&\bigg(\frac{f-1}{\upsilon
(1+\omega )}\bigg)^{b_1}\bigg(\frac{4 b_1(\psi  - 1)}{ b_1-1
}\bigg),\\\nonumber b_5&=&\bigg(\frac{M_4^2A f \lambda }{4\pi
}\bigg) \bigg(\frac{1}{2(1+b)}\bigg)^{2b_1}\bigg(\frac{\xi_1
}{2b_1-1}\bigg).
\end{eqnarray}
Figure \textbf{3} shows that the value of spectral index $n_s$ is
compatible with the number of e-folds, which are approximately equal
to $30$. According to WMAP$7$ \cite{b1, b2}, WMAP$9$ \cite{b4} and
Plank $2015$ \cite{b5}, the value of spectral index lies in ranges
$0.967 \pm 0.014$, $0.972 \pm 0.013$, and $ 0.968 \pm 0.006$.
\begin{figure}
\centering \epsfig{file=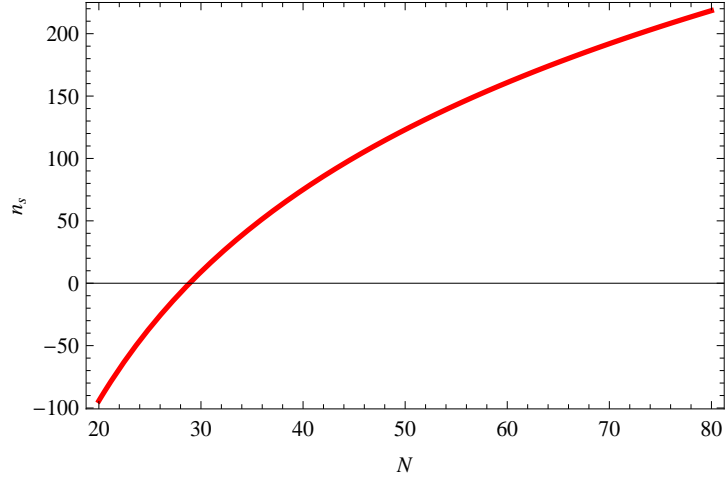, width=.70\linewidth, height=2.5in}
\caption{Graph of Spectral index for in term of e-folds for $\lambda
= 0.5, \sigma =1, A = 4, l= 1, f =\frac{3}{5}, \psi =\frac{4}{3},
\xi_1 =0.2\times10^{-8}, \omega  = 0.25, \upsilon = 1$.}
\end{figure}

The running of spectral index can be defined as follows \cite{b7}
\begin{eqnarray}\nonumber
\alpha_s = \frac{d n_s}{d\ln{K}}.
\end{eqnarray}
By using expression of $n_s$, $d\ln{K}= - d{N}$ and Eq.(\ref{phif}),
$\alpha_s$ leads to
\begin{eqnarray}\nonumber
\alpha_s&=&\frac{2^{3+4 b_1}}{(A f)^2}
\bigg(\frac{1}{1+b}\bigg)^{1-4 b_1} \exp\bigg[-\phi ^{4 b_1-2} b_3
b_5\bigg] \phi ^{-4 (2+b_1)} \bigg(\frac{-1+f}{\upsilon  (1+\omega
)}\bigg)^{-2 b_1} b_1^2
\\\nonumber&\times& \bigg[(-11 (1+b) \exp\bigg[\phi ^{2 b_1} b_3 b_5\bigg]+4
\exp\bigg[\phi ^{-2+4 b_1} b_3 b_5\bigg] (5f-7)) \phi ^4
(1+b_1)\\\nonumber&-&8 (1+b) \exp\bigg[\phi ^{2 b_1} b_3 b_5\bigg]
\phi ^{6 b_1} (-1+2 b_1) b_3^2 b_5 ((1-b_1) b_4+\phi ^2 b_1
b_5)\\\nonumber&+&(1+b) \exp\bigg[\phi ^{2 b_1} b_3 b_5\bigg] \phi
^{2+2 b_1} b_3 (16 (b_1-1) b_4+3 \phi ^2 b_1 b_5)+(1+b)
\\\nonumber&\times&\exp\bigg[\phi ^{2 b_1} b_3 b_5\bigg] \phi ^{2+4
b_1} b_3 b_5 \bigg(11+2 b_1 (-11+4 b_3 ((1-b_1) b_4+\phi ^2 b_1
b_5))\bigg)\bigg].
\end{eqnarray}
\begin{figure}
\centering \epsfig{file=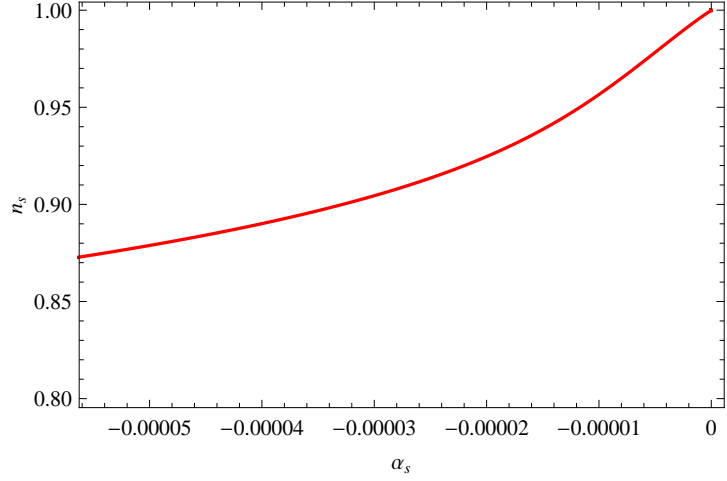, width=.70\linewidth, height=2.5in}
\caption{Plot of spectral index with it's running for $\lambda =0.2,
\sigma =1, A =1, M_4 = 1, f =\frac{3}{7}, \psi =\frac{4}{3}, \xi_1 =
0.2\times 10^{-8}, T = 5.47\times 10^{-5}, \omega  = 0.25, \upsilon
= 1$.}
\end{figure}

Figure \textbf{4} represents that spectral index ($n_s$) versus
running of spectral index is compatible with both observational
data. For example, WMAP$7$ observational data has provided the value
of spectral index and it's running which are approximately equal to
$1.027\pm0.051$ and $-0.034\pm0.026$, respectively. According to
WMAP$7$ \cite{b1,b2} and WMAP$9$ \cite{b4}, the value of spectral
index and it's running are approximately equal to $1.009\pm0.049,
0.992 \pm 0.019, $ and $ -0.019\pm0.025, -0.019 \pm 0.025$
respectively.

\begin{figure}
\centering \epsfig{file=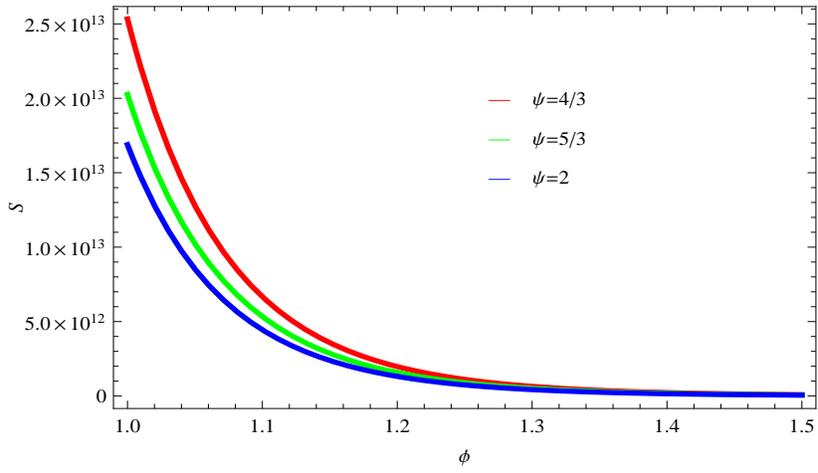, width=.80\linewidth,
height=2.5in} \caption{Plot of entropy density $S$ in terms of
inflaton $\phi$ use the parametric values as $\lambda = 1,\sigma =1,
A = 1, M_4= 1, f = \frac{3}{5}, \Gamma_1  = 0.275, \xi_2 = 0.2
\times 10^{-8}, T =5.47\times 10^{-5}, \omega  =0.25, \beta = 0.775,
\upsilon = 1$.}
\end{figure}

\subsection{Constant Bulk and Dissipative Coefficients}

Here, we take $\Gamma=\Gamma_1$, $\xi=\xi_2$. Under these
considerations, $\phi(t)$ and $V(\phi)$ lead to
\begin{eqnarray}\nonumber
\phi -\phi _0&=&\frac{2 }{1+h}\bigg(\frac{3\lambda  M_4^2}{4 \pi
}\bigg)^{\frac{1+\sigma }{4}}\bigg(\frac{1-f}{\upsilon \Gamma
(1+\omega )}\bigg)^{\frac{1}{2}} t^{\frac{1+h}{2}}, \\\nonumber
V&=&(A
f)^2\bigg(\frac{1+h}{2}\bigg)^{\frac{4(f-1)}{1+h}}\bigg(\frac{4 \pi
}{3 \lambda  l^2}\bigg)^{\frac{(1+\sigma
)(f-1)}{(1+h)}-1}\bigg(\frac{\upsilon  \Gamma_1 (1+\omega
)}{1-f}\bigg)^{\frac{2(f-1)}{1+h}}\phi ^{\frac{4(f-1)}{1+h}},
\end{eqnarray}
where $h =(f-1)(1+ \sigma )$. In this case, entropy density reduces
to
\begin{eqnarray}\nonumber
S&=&\bigg(\frac{A f}{T \psi}\bigg)\bigg(\frac{1+h}{2}\bigg)^{\frac{2
(f-1)}{1+h}}\bigg(\frac{4 \pi }{3 \lambda
M_4^2}\bigg)^{\frac{(1+\sigma )(f-1)}{2(1+h)}-1}\bigg(\frac{\upsilon
\Gamma_1 (1+\omega )}{1-f}\bigg)^{\frac{(f-1)}{1+h}}\phi
^{\frac{2(f-1)}{1+h}}\\\nonumber&\times&\bigg[\frac{(A f)^2}{
\Gamma_1 }\bigg(\frac{4
(f-1)}{1+h}\bigg)^2\bigg(\frac{1+h}{2}\bigg)^{\frac{4(f-1)}{1+h}}\bigg(\frac{4
\pi }{3\lambda  M_4^2}\bigg)^{\frac{(1+\sigma )(f-1)}{(1+h)}}\phi
^{\frac{4(f-1)}{1+h}-2}\\\nonumber&\times&\bigg(\frac{\upsilon
\Gamma_1 (1+\omega )}{1-f}\bigg)^{\frac{2(f-1)}{1+h}}+ \xi_2 \bigg].
\end{eqnarray}
Its plot versus $\phi$ is shown in Figure \textbf{5} which also
remains positive as well as exhibits the decreasing behavior for
three different values of $\psi$. The slow-roll parameter $\epsilon$
is obtained for this case is
\begin{eqnarray}\nonumber
\epsilon = \frac{1-f}{A f}\bigg(\frac{1+h}{2}\bigg)^{\frac{-2f
}{1+h}}\bigg(\frac{4 \pi }{3\lambda M_4^2}\bigg)^{\frac{-(1+\sigma
)f}{2(1+h)}}\bigg(\frac{\upsilon \Gamma_1 (1+\omega
)}{1-f}\bigg)^{\frac{-f}{1+h}}\phi ^{\frac{-2f}{1+h}}.
\end{eqnarray}
For this case, the lower and upper bounds of $\phi$ can be obtained
by using $\epsilon\simeq 1$ as
\begin{eqnarray}\nonumber
\phi _i &=& \bigg(\frac{1-f}{A
f}\bigg)^{\frac{1+h}{2f}}\bigg(\frac{2}{1+h}\bigg)\bigg(\frac{3\lambda
M_4^2}{4 \pi }\bigg)^{\frac{1+\sigma }{4}}\bigg(\frac{1-f}{\upsilon
\Gamma_1 (1+\omega )}\bigg)^{\frac{1}{2}},\\\nonumber \phi _g
&=&\bigg(\frac{2}{1+h}\bigg)\bigg(\frac{3\lambda  M_4^2}{4 \pi
}\bigg)^{\frac{1+\sigma }{4}}\bigg(\frac{1-f}{\upsilon  \Gamma_1
(1+\omega )}\bigg)^{\frac{1}{2}}\bigg(\frac{N}{A}+\bigg(\frac{1-f}{A
f}\bigg)\bigg)^{\frac{1+h}{2f}}.
\end{eqnarray}
The auxiliary function turns out to be
\begin{eqnarray}\nonumber
\chi(\phi)&=&\frac{3}{2}\bigg(\frac{1-f}{1+h}\bigg)\ln\phi.
\end{eqnarray}
The scalar power spectra and corresponding amplitude in this case
becomes
\begin{eqnarray}\nonumber
P_r&=&\frac{(A f)^{\frac{3}{2}}T_r}{2\sqrt{\Gamma_1}\pi^2(1-f)}
\bigg(\frac{1+h}{2}\bigg)^{\frac{5-3f}{1+h}}\bigg(\frac{4 \pi
}{3\lambda  M_4^2}\bigg)^{\frac{2(1+\sigma
)f-5h+6}{4(1+h)}}\bigg(\frac{\upsilon  \Gamma_1 (1+\omega
)}{1-f}\bigg)^{\frac{5-3f}{2(1+h)}}\\\nonumber&\times&\phi
^{\frac{-2}{1+h}},\\\nonumber A_g^2&=&\frac{(A f)^2}{2\pi
^2}\bigg(\frac{1+h}{2}\bigg)^{\frac{4(f-1)}{1+h}}\bigg(\frac{4 \pi
}{3\lambda  M_4^2}\bigg)^{\frac{h}{(1+h)}}\bigg(\frac{\upsilon
\Gamma_1 (1+\omega )}{1-f}\bigg)^{\frac{2(f-1)}{1+h}}\phi
^{\frac{4(f-1)}{1+h}}\\\nonumber&\times&\coth\bigg[\frac{K}{2
T}\bigg].
\end{eqnarray}
The tensor-to-scalar ratio becomes
\begin{eqnarray}\nonumber
r(K_0)&=&\frac{(A f)^4(1-f)}{T_r}\sqrt{\frac{\Gamma_1 }{A
f}}\bigg(\frac{1+h}{2}\bigg)^{\frac{7 f-9}{1+h}} \bigg(\frac{4 \pi
}{3\lambda  M_4^2}\bigg)^{\frac{9 h-6-2 f (1+\sigma )}{4 (1+h)}}\phi
^{\frac{4 f-2}{1+h}} \\\nonumber&\times&\bigg(\frac{\upsilon
\Gamma_1 (1+\omega )}{1-f}\bigg)^{\frac{7 f-9}{2
(1+h)}}\coth\bigg[\frac{K}{2 T}\bigg].
\end{eqnarray}
\begin{figure}
\centering \epsfig{file=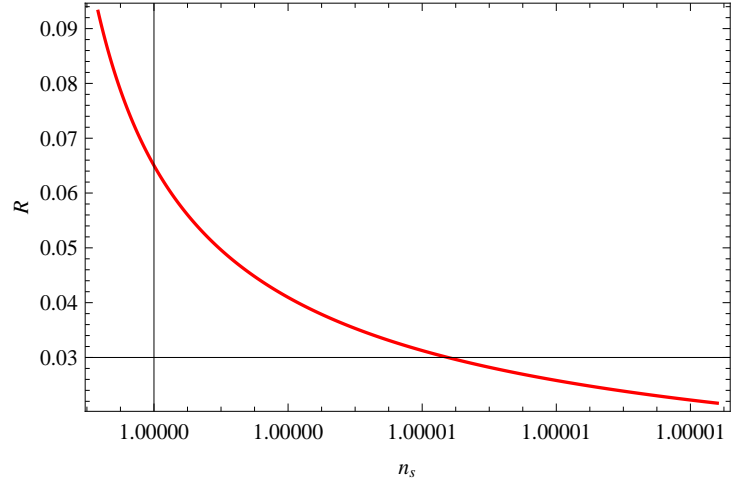, width=.70\linewidth, height=2.5in}
\caption{Plot for scalar-tensor ratio verses spectral index for
$\lambda = 1, \sigma =1, A = 5, M_4= 1, f =\frac{3}{5}, \psi
=\frac{4}{3}, T = 5.47\times 10^{-5}, \Gamma_1 =0.75, \omega  =
0.25, \beta = 0.775, P=2.541\times 10^4, K=0.002$.}
\end{figure}

The plot of this tensor-to-scalar ratio is shown in Figure
\textbf{6} and we obtain $r < 0.11$ in the present scenario.
However, the WMAP$7$, WMAP$9$ and Plank data \cite{b1, b2} predict
an upper bound for tensor-to-scalar ratio $r < 0.36, 0.38, 0.11$ for
spectral index $n_s = 0.982 \pm 0.020,~0.992 \pm 0.019,~ 0.9655 \pm
0.0062$ respectively.
The scalar spectral index leads to
\begin{eqnarray}\nonumber
n_s&=&1+\frac{\Gamma_1 }{2 (A f)^2 (1-f)
}\bigg(\frac{1+h}{2}\bigg)^{\frac{6(1- f)}{1+h}} \bigg(\frac{4 \pi
}{3\lambda  M_4^2}\bigg)^{1-\frac{3 h}{2 (1+h)}}
\bigg(\frac{\upsilon  \Gamma_1 (1+\omega )}{1-f}\bigg)^{\frac{3(1-
f)}{1+h}}\\\nonumber&\times&\phi ^{\frac{6(1- f)}{1+h}}.
\end{eqnarray}
\begin{figure}
\centering \epsfig{file=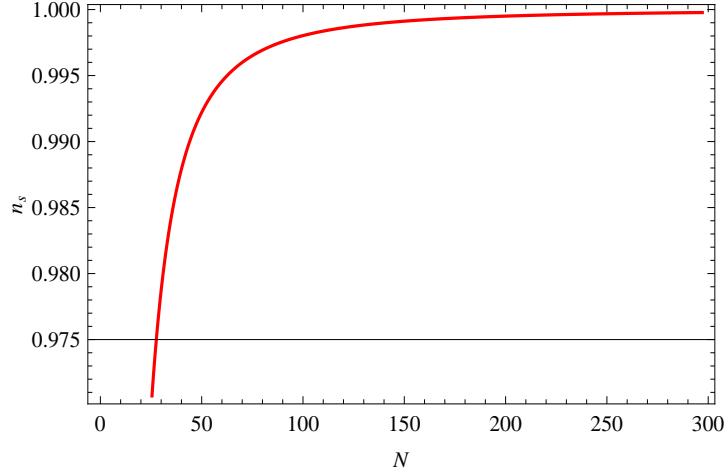, width=.70\linewidth, height=2.5in}
\caption{Plot of Spectral index in term of e-folds for $\lambda=
0.2, \sigma =1, A = 3, M_4= 1, f =\frac{3}{5}, T = 5.47\times
10^{-5}, \Gamma_1 =0.275, \omega  = 0.25, \upsilon = 1$.}
\end{figure}
Figure \textbf{7} shows that the value of spectral index $0.96 < n_s
< 0.97$ is compatible with the number of e-folds, which are
approximately equal to $30$ (WMAP$7$ \cite{b1, b2}, WMAP$9$
\cite{b4} and Plank 2015 \cite{b5}). The $\alpha_s$ takes the
following form
\begin{eqnarray}\nonumber
\alpha_s&=&\frac{3\Gamma _1^2}{4(f-1)(A
f)^5}\bigg(\frac{1+h}{2}\bigg)^{\frac{12 (1-f)}{1+h}} \bigg(\frac{4
\pi }{3\lambda M_4^2}\bigg)^{-1+\frac{3}{1+h}} \bigg(\frac{\upsilon
\Gamma_1 (1+\omega )}{1-f}\bigg)^{\frac{6(1-
f)}{1+h}}\\\nonumber&\times& \phi ^{\frac{12 (1-f)}{1+h}}.
\end{eqnarray}
\begin{figure}
\centering \epsfig{file=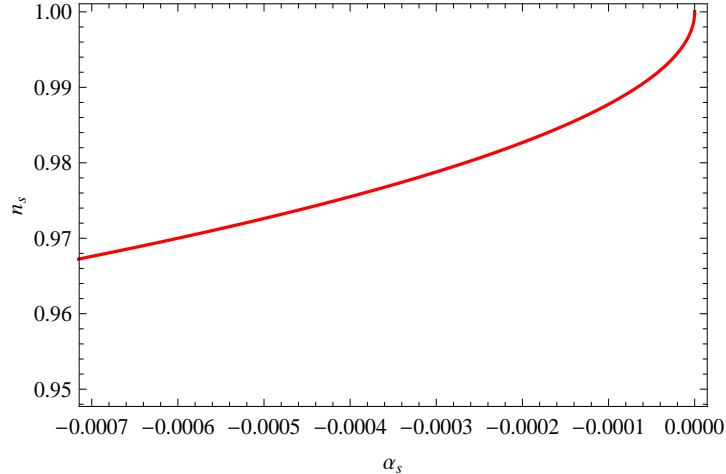, width=.70\linewidth, height=2.5in}
\caption{Plot of spectral index with it's running for $\lambda= 0.5,
\sigma =1, A = 3, M_4= 1, f =\frac{3}{5}, \psi =\frac{4}{3}, \xi_2 =
3.2 \times 10^{-8}, T = 5.47\times 10^{-5}, \omega  = 0.25, \Gamma_1
=0.275, \beta = 0.775, \upsilon = 1$.}
\end{figure}
Figure \textbf{8} represents that running of spectral index is
compatible with both observational data (WMAP $7+9$) for constant
bulk and dissipative coefficients.

\section{Conclusion}

We have considered the warm intermediate MCG inflationary scenario
on the brane with bulk viscous pressure and high dissipative regime
in flat FRW universe. We have calculated the slow-roll parameters,
number of e-folds, scalar-tensor power spectra, spectral indices,
tensor–scalar ratio and running of scalar spectral index. We have
analyzed these parameters for variable as well as constant
dissipation and bulk viscous coefficients. We have restricted
constant parameters involving in the models according to WMAP$7$
results for examining the physical behavior of $n_s - N$, $n_s - r$
and $n_s - \alpha_s$ trajectories in both cases of dissipation and
bulk viscous coefficients. We have chosen model parameters of MCG as
$\omega=0.25,~\sigma=1,~\beta=0.775$ which lies within the
constraints $(-1.186,0.2754), ~(-0.9469,1.442)$ and
$(0.0497,0.9935)$ obtained by \cite{paul}, respectively.

The entropy density has been displayed versus scalar field ($\phi$)
(Figure \textbf{1} and \textbf{5}) which shows increasing behavior
and remains positive (as expected) for both cases of dissipation and
bulk viscous coefficients. The standard values of tensor-scalar
ratio $r<0.36,~ 0.38,~ 0.11$ for spectral index $n_s=0.982\pm 0.020,~
0.992\pm0.019,~ 0.9655\pm0.0062$ according to WMAP$7$ \cite{b1, b2},
WMAP$9$ \cite{b4} and Plank 2015 \cite{b5} result respectively. In
our case, the tensor-scalar ratio versus spectral index is
compatible with this observational data (Figure \textbf{2} and
\textbf{6}). We have also observed from Figures (\textbf{3} and
\textbf{7}) that the trajectories of spectral index lies within the
suggested ranges of observations for approximately $30$ number of
e-folds, i.e., $n_s=1.027\pm0.051,~ 1.009\pm0.049,~ 0.096\pm0.025$
and $\alpha_{s}=-0.034\pm0.026,~ -0.019\pm0.025,~ -0.0084\pm0.0082$
according to WMAP$7$ \cite{b1, b2}, WMAP$9$ \cite{b4} and Plank 2015
\cite{b5}, respectively. We have also attained the compatibility of
spectral index and it's running with the before mentioned
observational schemes (Figures \textbf{4} and \textbf{8}).

In \cite{m1}, we have examined the possible realization of warm
chaotic inflation and logamediate inflation within the framework of
a MCG brane-world model by assuming the standard scalar field. We
have investigated the slow-roll parameters, number of e-folds,
scalar-tensor power spectra, spectral indices, tensor–scalar ratio
and running of scalar spectral index for variable as well as
constant dissipation and bulk viscous coefficients. We have found
$r< 0.11$, $n_s=0.96\pm0.025$ and $\alpha_s =-0.019\pm0.025$ and
these ranges are consistent with BICEP$2$, WMAP $(7+9)$ and Planck
data \cite{m1}. Also, we have investigated the MCG inspired
inflationary regime in the brane-world framework in the presence of
standard and tachyon scalar fields \cite{m}.

We have developed the $n_s - N$ and $r - N$ planes and concluded
that $n_s\simeq96^{+0.5}_{-0.5}$ and $r\leq0.0016$ for
$N\simeq60^{+5}_{-5}$ in both cases of scalar field models as well
as for all values of $m$ and these constraints are consistent with
observational data such as WMAP$7$, WMAP$9$ and Planck data
\cite{m}. We have also explored warm inflationary universe models by
assuming with various chaplygin gas models with $\Gamma\propto T$,
weak and strong dissipative regimes and quartic potential
$\frac{\lambda_{*}\phi^{4}}{4}$ \cite{m3}. We have observed that the
$r<0.05$ in generalized chaplygin gas, $r<0.15$ in modified
chaplygin gas, and $r<0.12$ in generalized cosmic chaplygin gas
models and these are in agreement with WMAP$9$ and latest Planck
data.

The present work is different from \cite{m1,m,m3}. We have explored
the role of bulk viscous pressure on the warm intermediate
inflationary MCG in brane-world framework by taking standard scalar
field. We have found that the behavior of entropy density in the
current scenario ensures the validity of thermodynamics laws. We
have illustrated the results $n_s - N$, $n_s - r$ and $n_s -
\alpha_s$ for variable as well as constant dissipation and bulk
viscous coefficients at high dissipative regime and found that these
parameters are compatible with recent observational data such as
WMAP $7+9$, BICEP$2$ and Planck data.

\end{document}